# Data Science: A Powerful Catalyst for Cross-Sector Collaborations to Transform the Future of Global Health - Developing a New Interactive Relational Mapping Tool (Demo)

Barbara Bulc
Global Development
Geneva, Switzerland
bbulc@gd-impact.org

Cassie Landers
Columbia University
Mailman School of Public Health
New York, NY, USA
cl689@cumc.columbia.edu

Katherine Driscoll
Columbia University
Mailman School of Public Health
New York, NY, USA
ked2166@cumc.columbia.edu

## ABSTRACT
The increasingly complex and rapidly changing global health and socio-economic landscape requires fundamentally new ways of thinking, acting and collaborating to solve growing systems challenges. Cross-sectoral collaborations between governments, businesses, international organizations, private investors, academia and non-profits are essential for lasting success in achieving the Sustainable Development Goals (SDGs), and securing a prosperous future for the health and wellbeing of all people.[1] Our aim is to use data science and innovative technologies to map diverse stakeholders and their initiatives around SDGs and specific health targets - with particular focus on SDG 3 (Good Health & Well Being) and SDG 17 (Partnerships for the Goals) - to accelerate cross-sector collaborations. Initially, the mapping tool focuses on Geneva, Switzerland as the world center of global health diplomacy with over 80 key stakeholders and influencers present. As we develop the next level pilot, we aim to build on users' interests, with a potential focus on non-communicable diseases (NCDs) as one of the emerging and most pressing global health issues that requires new collaborative approaches. Building on this pilot, we can later expand beyond only SDG 3 to other SDGs.

## NO ORGANIZATION CAN SOLVE COMPLEX CHALLENGES ALONE
The rapidly growing burden of NCDs looms with over 75% of deaths occurring in the developing countries, becoming only more complex as the world's population ages, our lifestyles change and environmental challenges grow. According to the World Health Organization, worldwide obesity has more than doubled over the past two decades, causing increased rates of heart disease and stroke especially in low- and middle-income countries.[2] Although it is entirely preventable, curbing global obesity requires a population-based, multi-sectoral, multi-disciplinary, and culturally relevant approach.[3] Additionally, what some may consider diseases of the past, like tuberculosis (TB) or cholera, are still devastating killers. We know the solutions but these diseases remain rampant, in large part due to the lack of integrated approaches and effective collaborations. According to the Stop TB Partnership nearly 1.5 million people lose their lives each year to TB, making it today, 135 years after the discovery of the bacterium, still the leading global infectious killer.[4] Accelerated multi-sectoral coordination and action involving health and other sectors is the only way of stopping TB.[5]

**RAPIDLY CHANGING ENVIRONMENT**
At the same time, shifting socio-economic environment and increasing health expenditures require innovative financing solutions and new collaborations among public and private stakeholders if we are to achieve universal health coverage (UHC). For the first time in history, more than half of the world's 100 largest economies are companies, not countries, and cities are making and implementing policies.[6] Businesses from all industry sectors have a critical role to play.[7] United Nations agencies have overlapping mandates and are slow to act without effective partnerships. There is increasing number of various non-profits, looking for new collaborations and competing for resources. The emergence of organized citizens' voices and coalitions is evermore powerful in shaping our future.

No single organization is equipped to solve increasing global health challenges alone. Over the past few decades, a number of groundbreaking public –private partnerships or alliances like The Global Fund to Fight AIDS, Tuberculosis and Malaria, UNITAID, Gavi, the Vaccine Alliance, Roll Back Malaria, Stop TB Partnership, Global Alliance for Improve Nutrition and several others have been developed to tackle global health problems. Nevertheless, solutions and financing have been framed around specific diseases or sectors causing silos to emerge. How can we tap into the capabilities and resources of various public and private stakeholders using data science and technology to bridge these silos and build a more effective collaborative system, accelerating the delivery of sustainable solutions?

## A NEW TOOL TO HELP NAVIGATE PARTNERSHIP OPPORTUNITIES
In this current era of technological revolution, as some call it the Fourth Industrial Revolution,[8] we have a unique opportunity to use technologies for finding the best strategic partners and accelerating effective collaborations among diverse stakeholders and sectors, globally and locally. It is perhaps surprising that, in our interconnected world, technologies have not been extensively used to catalyze and cultivate strategic relationships between organizations.

Global Development (GD), a new kind of consultancy focused on building collaborative ventures to address health systems challenges and sustainable development, partnered with Columbia University's Mailman School of Public Health and Kumu, a data visualization platform startup, to conceptualize and develop an interactive, visual tool of global health stakeholders. Our aim was to create a tool for the public good and as an open data platform, to foster and accelerate the development of cross-sector partnerships, which enable bringing sustainable solutions to scale.

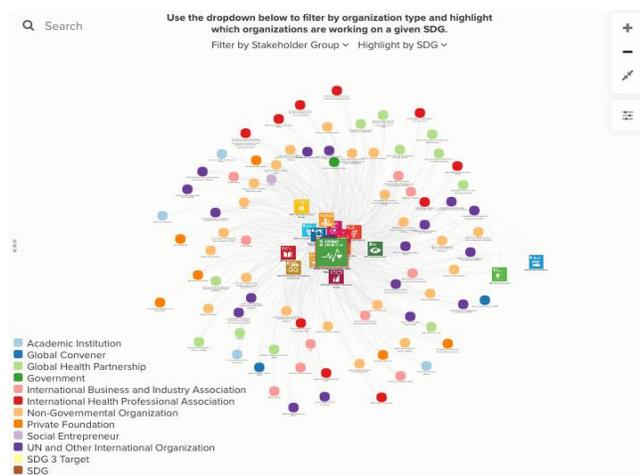

FIGURE 1: An interactive relational map of key public and private actors in global health in Geneva, Switzerland, the world center of global health diplomacy

As the launchpad for our pilot we chose Geneva - the world center for global health and multilateral diplomacy, which hosts likely the highest number of key influencers in global policy making, financing and implementation of various programs and initiatives in global health.

We developed a relational database and "sense-making" system map of over 84 global health stakeholders based in Geneva. Stakeholders were sorted in the following ten categories (listed alphabetically): academia, global convener, global health partnership, government, international business and industry association, health professional association, non-governmental organization, private foundation, social entrepreneurship, and the United Nations and other international organizations. Each stakeholder's activities were mapped across nine targets defined in SDG 3 (Good Health & Wellbeing), as well as their focus on other SDGs (1-17). Additionally, we included in the database their mission, annual budget, staff size, headquarter location and regions of focus. For a more visually appealing and clearer look, we represented organizations by their logo, and SDGs by their formal symbols.

The organizations in the mapping tool were elaborated and finalized using both online research and email inquiries sent to each organization. In addition, in-person interviews were conducted with a number of organizations for qualitative feedback on making the map useful and user-friendly.

This multi-stakeholder map, displayed relationally, is only a first step to help organizations and individuals identify synergistic partners and visualize how to best facilitate new collaborations between stakeholders of all types. It enables bridging silos in a highly fragmented global health landscape.

## FUTURE VISION TO ACCELERATE COLLABORATIONS

This initial prototype generated excitement and positive response from its participants and users. It addresses the present lack of awareness of innumerable private and public sector actors and their extensive activities. It also helps to visualize the interconnected nature of health with other issues such as environment and climate change, and many opportunities in health and beyond. Several users, including students of global health and sustainable development, found it to be an effective and much needed educational tool.

**SUGGESTIONS FOR NEXT LEVEL PROTOTYPE**

The development of the next level prototype is underway. Suggestions for future development included developing different use-cases for specific topics and geographies, making the tool more transactional and open source, enabling the development of a more collaborative community. This would require testing the prototype and making it more robust and user friendly.

In addition, it would be useful to explore the integration of text-mining and smart-classification to speed up research and extraction of relevant data from online sources, while also adopting the use of a more relevant type of database (NoSQL, e.g. MongoDB, OrientDB etc.) to represent more complex relationships. Early discussion is underway to standardize and publish an open ontology/taxonomy for representing data of this type, that could be queried and utilized by other consumers of data (e.g. as part of the open data framework).

Based on received feedback, NCDs emerged as one of the highly desired future directions. Specifically, mapping key actors, their projects and investments in select low- and middle-income countries, or cities, to help drive collective impact at scale in this currently highly fragmented area. Another expressed potential area of interest is to adapt the tool for various membership-based coalitions, such as the Global Health Council, NCD Alliance, Stop TB Partnership, Partnership for Maternal, Newborn and Child Health and other global, regional, or local alliances in global health.

## MOVING BEYOND GLOBAL HEALTH

Furthermore, moving beyond the global health space, this tool can be used to help build global collaborations across other interrelated areas of the SDGs such as education, clean water and sanitation, food security, clean energy, gender equalities and others. Along with suggestions and enthusiasm received from our users, we are open and excited to further develop strategic collaborations and significantly improve and test this demo.

## CONCLUSION

For us as human beings relationships are at the core of our wellbeing and prosperity. Like never before, the opportunity to use data science and human capital in novel ways to enhance relationships among organizations, can help accelerate solving the challenges that we face in global health and beyond.

## ACKNOWLEDGEMENTS

We gratefully acknowledge the contributions of time, resources and insights from the following individuals and organizations: Cassie Landers, Assistant Professor of Population and Family Health and Faculty Lead, Office of Field Practice; Ana Jiménez-Bautista, Director
Office of Field Practice at Columbia University Mailman School of Public Health; Carol Liang, Caitlin Zuehlke, Jinjin Wu, Julie Kvedar, Master of Public Health students, Columbia University Mailman School of Public Health and Jeff Mohr, Cofounder and CEO of Kumu. We would specially like to thank David Hagan, International Consultant in Tech4Dev for his insights into the future direction of this tool and methodology. Support and enthusiasm from numerous participating organizations and interviewed individuals allowed us to explore more fully their needs, relational challenges and how to address them.